\let\ifarxiv=\iftrue     
\pdfoutput=1
{\def\usepackage{ws-procs9x6}}

\ifarxiv

\documentclass[12pt,a4paper]{article}

\ifnum\pdfoutput=1\else
\PassOptionsToPackage{hypertex}{hyperref}
\PassOptionsToPackage{draft}{graphicx}
\usepackage{showkeys}
\fi

\usepackage{amsmath,amssymb}
\usepackage[bookmarks=true,hyperfigures=true]{hyperref}
\usepackage{graphicx}
\usepackage[nosort]{cite}
\usepackage[bulletsep]{collref}

\usepackage[a4paper,text={450pt,650pt},centering]{geometry}

\usepackage[font=small,labelfont=bf,width=0.85\textwidth]{caption}

\numberwithin{equation}{section}

\else

\ifnum\pdfoutput=1\else
\PassOptionsToPackage{hypertex}{hyperref}
\PassOptionsToPackage{draft}{graphicx}
\PassOptionsToClass{draft}{ws-procs9x6}
\fi

\documentclass{ws-procs9x6}

\usepackage[bookmarks=true,hyperfigures=true]{hyperref}
\usepackage{graphicx}
\fi

\let\oldbfseries=\bfseries
\let\oldmdseries=\mdseries
\let\oldnormalfont=\normalfont
\renewcommand{\bfseries}{\oldbfseries\boldmath}
\renewcommand{\mdseries}{\oldmdseries\unboldmath}
\renewcommand{\normalfont}{\oldnormalfont\unboldmath}

\allowdisplaybreaks[3]

\providecommand{\hypersetup}[1]{}

\providecommand{\texorpdfstring}[2]{#1}
\providecommand{\pdfbookmark}[3][]{}

\hypersetup{plainpages=false}
\hypersetup{pdfpagemode=UseNone}
\hypersetup{bookmarksnumbered=true}
\hypersetup{pdfstartview=FitH}
\hypersetup{colorlinks=false}
\hypersetup{citebordercolor={.5 1 .5}}
\hypersetup{urlbordercolor={.5 1 1}}
\hypersetup{linkbordercolor={1 .7 .7}}

\DeclareMathSymbol{\Gamma}{\mathalpha}{letters}{"00}
\DeclareMathSymbol{\Delta}{\mathalpha}{letters}{"01}
\DeclareMathSymbol{\Theta}{\mathalpha}{letters}{"02}
\DeclareMathSymbol{\Lambda}{\mathalpha}{letters}{"03}
\DeclareMathSymbol{\Xi}{\mathalpha}{letters}{"04}
\DeclareMathSymbol{\Pi}{\mathalpha}{letters}{"05}
\DeclareMathSymbol{\Sigma}{\mathalpha}{letters}{"06}
\DeclareMathSymbol{\Upsilon}{\mathalpha}{letters}{"07}
\DeclareMathSymbol{\Phi}{\mathalpha}{letters}{"08}
\DeclareMathSymbol{\Psi}{\mathalpha}{letters}{"09}
\DeclareMathSymbol{\Omega}{\mathalpha}{letters}{"0A}

\newcommand{\gen}[1]{\mathrm{#1}}
\newcommand{\genY}[1]{\widehat{\mathrm{#1}}}
\newcommand{\superN}{\mathcal{N}}
\newcommand{\copro}{\mathrm{\Delta}}

\newcommand{\alg}[1]{\mathfrak{#1}}
\newcommand{\grp}[1]{\mathrm{#1}}
\newcommand{\yangian}{\mathrm{Y}}
\newcommand{\ham}{\mathrm{\hat{H}}}
\newcommand{\proj}{\mathrm{\hat{P}}}

\newcommand{\osc}[1]{\mathrm{#1}}

\newcommand{\Tr}{\mathop{\mathrm{Tr}}}
\newcommand{\sign}{\mathop{\mathrm{sign}}}

\newcommand{\order}[1]{\mathcal{O}(#1)}
\newcommand{\Op}{\mathcal{O}}
\newcommand{\Amp}{\mathcal{A}}
\newcommand{\cder}{\mathcal{D}}
\newcommand{\fld}{\mathcal{W}}
\newcommand{\fldscl}{\mathcal{B}}
\newcommand{\fldferm}{\mathcal{C}}

\newcommand{\fldstr}{\mathcal{F}}

\newcommand{\Integers}{\mathbb{Z}}
\newcommand{\Complex}{\mathbb{C}}

\newcommand{\gammafn}{\mathrm{\Gamma}}
\newcommand{\psifn}{\mathrm{\Psi}}

\ifx\genfrac\sdflkaj\else\fi
\newcommand{\sfrac}[2]{{\textstyle\frac{#1}{#2}}}
\newcommand{\half}{\sfrac{1}{2}}

\newcommand{\quarter}{\sfrac{1}{4}}

\newcommand{\indup}[1]{_{\mathrm{#1}}}

\newcommand{\supup}[1]{^{\mathrm{#1}}}

\newcommand{\lrbrk}[1]{\left(#1\right)}
\newcommand{\bigbrk}[1]{\bigl(#1\bigr)}

\newcommand{\vev}[1]{\langle#1\rangle}

\newcommand{\bigvev}[1]{\bigl\langle#1\bigr\rangle}
\newcommand{\bigcomm}[2]{\big[#1,#2\big]}
\newcommand{\comm}[2]{[#1,#2]}

\newcommand{\acomm}[2]{\{#1,#2\}}

\newcommand{\set}[1]{\{#1\}}
\newcommand{\state}[1]{\mathopen{|}#1\mathclose{\rangle}}

\newcommand{\sprod}[2]{\langle#1,#2\rangle}
\newcommand{\sprods}[2]{\langle#1#2\rangle}
\newcommand{\cprod}[2]{[#1,#2]}

\newcommand{\nln}{\nonumber\\}
\newcommand{\nl}[1][0pt]{\nonumber\\[#1]&\hspace{-4\arraycolsep}&\mathord{}}

\newcommand{\earel}[1]{\mathrel{}&\hspace{-2\arraycolsep}#1\hspace{-2\arraycolsep}&\mathrel{}}
\newcommand{\eq}{\earel{=}}
\newcommand{\beq}{\begin{equation}}
\newcommand{\eeq}{\end{equation}}

\def\[{\begin{equation}}
\def\]{\end{equation}}
\def\<{\begin{eqnarray}}
\def\>{\end{eqnarray}}

\makeatletter
\def\mr@ignsp#1 {\ifx\:#1\@empty\else #1\expandafter\mr@ignsp\fi}%
\newcommand{\multiref}[1]{\begingroup
\xdef\mr@no@sparg{\expandafter\mr@ignsp#1 \: }%
\def\mr@comma{}%
\@for\mr@refs:=\mr@no@sparg\do{\mr@comma\def\mr@comma{,}\ref{\mr@refs}}%
\endgroup}
\makeatother

\newcommand{\hypref}[2]{\ifx\href\asklfhas #2\else\href{#1}{#2}\fi}

\newcommand{\secref}[1]{Sec.~\multiref{#1}}

\newcommand{\figref}[1]{Fig.~\multiref{#1}}
\renewcommand{\eqref}[1]{(\multiref{#1})}

\makeatletter
\newlength{\apb@width}
\newcommand{\autoparbox}[2][c]{\settowidth{\apb@width}{#2}\parbox[#1]{\apb@width}{#2}}
\newcommand{\includegraphicsbox}[2][]{\autoparbox{\includegraphics[#1]{#2}}}
\makeatother

\providecommand{\href}[2]{#2}
\newcommand{\arxivlink}[1]{\href{http://arxiv.org/abs/#1}{arxiv:#1}}

\begin{document}

\pdfbookmark[1]{Title Page}{title}

\ifarxiv

\thispagestyle{empty}
\begin{flushright}\footnotesize
\texttt{\arxivlink{1004.5423}}\\
\texttt{AEI-2010-029}
\end{flushright}
\vspace{0.5cm}

\begin{center}%
{\Large\bfseries%
\hypersetup{pdftitle={On Yangian Symmetry in Planar N=4 SYM}}%
\hypersetup{pdfkeywords={Integrability, Yang-Mills, Local Operators, Anomalous Dimensions, Scattering Matrix}}%
\hypersetup{pdfsubject={To Lev Lipatov on the occasion of his 70th birthday}}%
On Yangian Symmetry in Planar $\superN=4$ SYM%
\par} \vspace{1cm}%

\textsc{Niklas Beisert}\vspace{5mm}%
\hypersetup{pdfauthor={Niklas Beisert}}%

\textit{Max-Planck-Institut f\"ur Gravitationsphysik\\%
Albert-Einstein-Institut\\%
Am M\"uhlenberg 1, 14476 Potsdam, Germany}\vspace{3mm}%

\texttt{nbeisert@aei.mpg.de}%
\par\vspace{1cm}

\textbf{Abstract}\vspace{7mm}

\begin{minipage}{12.7cm}
Planar $\mathcal{N}=4$ supersymmetric Yang--Mills theory
appears to be perturbatively integrable.
This work reviews integrability in terms of a Yangian algebra
and compares the application to the problems
of anomalous dimensions and scattering amplitudes.
\end{minipage}

\end{center}

\vspace{1cm}
\hrule height 0.75pt
\vspace{1cm}



\else

\title{\uppercase{On Yangian Symmetry in Planar $\mathcal{N}=4$ SYM}}
\hypersetup{pdftitle={On Yangian Symmetry in Planar N=4 SYM}}%

\author{\uppercase{Niklas Beisert}}
\hypersetup{pdfauthor={Niklas Beisert}}%

\address{Max-Planck-Institut f\"ur Gravitationsphysik\\%
Albert-Einstein-Institut\\%
Am M\"uhlenberg 1, 14476 Potsdam, Germany\\
E-mail: nbeisert@aei.mpg.de}

\begin{abstract}
Planar $\mathcal{N}=4$ supersymmetric Yang--Mills theory
appears to be perturbatively integrable.
This work reviews integrability in terms of a Yangian algebra
and compares the application to the problems
of anomalous dimensions and scattering amplitudes.
\end{abstract}

\hypersetup{pdfkeywords={Integrability, Yang-Mills, Local Operators, Anomalous Dimensions, Scattering Matrix}}%

\hypersetup{pdfsubject={Gribov-80 Memorial Volume: Quantum Chromodynamics and Beyond}}%

\bodymatter

\fi

\section{Introduction}

Integrability is a very useful feature of selected physical models. 
It allows one to rely on certain algebraic properties 
to solve them exactly and to determine physical observables efficiently. 
Unfortunately, in general integrability is restricted to 
at most two-dimensional models. 
These can be discrete, e.g.\ spin chains, statistical physics models, 
or continuous, e.g.\ sigma models such as
two-dimensional (super)gravity and worldsheet models string theory. 

Despite this severe restriction, signs of integrability have
been discovered in four-dimensional gauge theories: 
Lipatov noticed that the BFKL Hamiltonian \cite{Lipatov:1976zz,Kuraev:1976ge,Kuraev:1977fs,Fadin:1975cb,Balitsky:1978ic}
describing the evolution of reggeized gluons in QCD high-energy scattering 
is integrable and closely related to the Heisenberg spin chain \cite{Lipatov:1994xy,Faddeev:1994zg}
(see also \cite{Lipatov:2009nt} for a recent account and additional references). 
The crucial additional assumption which enables integrability in this 
four-dimensional model is the 't Hooft large-$N\indup{c}$ 
or \emph{planar limit} \cite{'tHooft:1974jz}.
In this limit the gauge group dynamics 
reduces to two-dimensional surfaces
on which the integrable structure lives.

Another instance of integrability in large-$N\indup{c}$ gauge theory 
is deep inelastic scattering where
anomalous dimensions of local operators 
are responsible for scaling violations.
The anomalous dimensions of local operators 
can be described by the DGLAP evolution equation which was 
initiated by Gribov and Lipatov \cite{Gribov:1972ri,Altarelli:1977zs,Dokshitzer:1977sg}.
It was noticed that also these evolution equations
are integrable to some extent \cite{Lipatov:1997vu,Braun:1998id,Lipatov:1998as,Braun:1999te,Belitsky:1999ru}.

In 2002 a new line of developments
started for a particular four-dimensional gauge theory, 
namely $\superN=4$ maximally supersymmetric Yang--Mills ($\superN=4$ SYM).
This model, consisting of a $\grp{U}(N\indup{c})$ gauge field,
4 flavours of massless adjoint fermions and 6 flavours of 
massless adjoint scalars, 
is relevant to the AdS/CFT string/gauge duality.
Integrability was shown to apply to 
all leading-order planar anomalous dimensions \cite{Minahan:2002ve,Beisert:2003tq}.
Unlike in the analogous problem in QCD, 
integrability was moreover demonstrated to survive in 
higher-order quantum corrections \cite{Beisert:2003yb,Beisert:2003ys}
hinting at complete integrability of the planar sector of the theory.

In this paper we review 
integrability of planar $\superN=4$ SYM 
in the guise of Yangian symmetry.
We shall focus on the problems of 
anomalous dimensions of local operators (\secref{sec:spinchains}) 
and the spacetime scattering matrix (\secref{sec:scattering})
in order to reveal the close similarities between them (\secref{sec:comparison}).

\section{Anomalous Dimensions of Local Operators}
\label{sec:spinchains}

Scaling dimensions of local operators represent a key set of 
observables in a conformal field theory. 
They determine to a large extent the 
spacetime dependence of correlation functions.
The spectrum of scaling dimensions 
can be viewed as the conformal analog of the mass spectrum of 
composite particles in a non-conformal field theory.
In $\superN=4$ SYM the planar spectrum turned out to be governed 
by an integrable system with an underlying Yangian algebra.
In the following we shall review local operators and the role
the Yangian algebra.

\subsection{Framework}

\begin{figure}
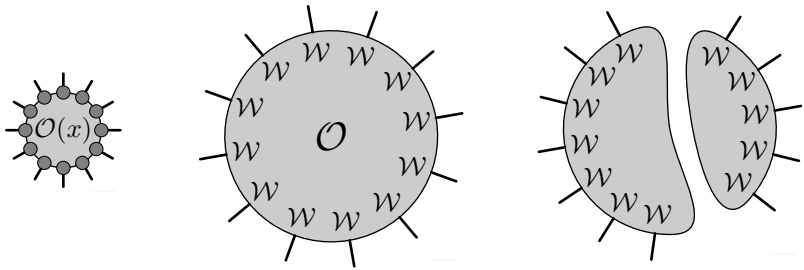
\centering
\includegraphicsbox[scale=\ifarxiv 1.0\else 0.8\fi]{FigChainTraceOp.mps}
\qquad
\includegraphicsbox[scale=\ifarxiv 1.0\else 0.8\fi]{FigChainTraceSingle.mps}
\qquad
\includegraphicsbox[scale=\ifarxiv 1.0\else 0.8\fi]{FigChainTraceDouble.mps}
\caption{Local operators as gauge invariant combination of fields
at a common point in spacetime (left). 
Focusing on the $\grp{U}(N\indup{c})$ gauge group structure alone, 
operators are classified by the number of traces: 
single-trace (middle) and multi-trace (right) operators.}
\label{fig:localops}
\end{figure}

Local operators are local, gauge-invariant combinations of the 
scalars $\fldscl$, fermions $\fldferm$ and gauge field strengths $\fldstr$
as well as their covariant derivatives $\cder$.
Gauge invariant combinations are constructed 
as traces of products of covariant fields, 
e.g.\
\<\label{eq:locopex}
\bar\Op_1(x)\eq \Tr \bigbrk{\fldscl_m(x)\,\fldscl_m(x)},
\nln
\bar\Op_2(x)\eq \Tr \bigbrk{\cder^\mu \fldscl_m(x) \,\cder_\mu \fldscl_n(x)},
\nln
\bar\Op_3(x)\eq\ldots.
\>
One can also construct multi-trace operators, such as $\bar\Op_1(x)\bar\Op_2(x)$, 
but in the planar limit these decouple and can be safely ignored,
see \figref{fig:localops}.

In a perturbative QFT on flat Minkowski spacetime 
the correlator of two such operators
takes the generic form 
$\vev{\bar\Op_A(x)\bar\Op_B(y)}=F_{AB}(x-y,g,\mu,\epsilon)$
due to Poincar\'e symmetry.
Here $\mu$ is the regularisation scale and 
$\epsilon$ is the parameter of dimensional regularisation.
Importantly, the result is generically divergent as one removes the
regulator, i.e.\ at $\epsilon\to 0$. 
This also applies to $\superN=4$ SYM.
Superficially, it contradicts the fact that $\superN=4$ SYM is a finite CFT 
where two-point functions take a particular form
which depends only on the \emph{scaling dimensions} $D_A$ 
of the local operators $\tilde\Op_A$
\footnote{This expression applies to scalar operators;
spinning operators have a different, yet uniquely determined
and $x$-dependent structure in the numerator.}
\[
\bigvev{\tilde\Op_A(x)\,\tilde\Op_B(y)}=\frac{\delta_{AB}}{|x-y|^{2D_A}}\,.
\]

To recover this form one has to find the right linear combinations
$\tilde\Op_A$ of the bare operators $\bar\Op_A$,
see e.g.\ \cite{Bianchi:1999ge}.
This is usually done in two steps:
First, renormalisation absorbs the divergencies
into the definition of the operators $\bar\Op_A=Z_A{}^B\Op_B$.
Then the operators are diagonalised to achieve the above form
by means of another linear transformation $\Op_A\to\tilde\Op_A$. 
Of course the composition of the two maps is yet another linear map, 
but it still makes sense to distinguish the two steps:
Renormalisation can be performed abstractly on 
a basis of states as in \eqref{eq:locopex} 
to the end that one can enumerate renormalised operators $\Op_B$ 
in an equivalent basis. 
Conversely, diagonalisation requires the precise knowledge 
of the set of operators one is interested in. 
Moreover it requires to solve algebraic equations, 
potentially of very high degree.
It should be noted though that the splitting remains
somewhat ambiguous to the extent that $Z$
is uniquely determined by the model 
only modulo some transformations
(which are eventually compensated by the diagonalisation).

Next we establish a useful basis of single-trace local operators
in $\superN=4$ SYM
\[\label{eq:locopgen}
\Op=\Tr \bigbrk{\fld_1 \fld_2 \fld_3 \ldots \fld_n},
\qquad
\fld_k\in \set{\cder^j\fldscl,\cder^j\fldferm,\cder^j\fldstr}.
\]
The matrices $\fld_k$ represent 
the scalars $\fldscl$, the fermions $\fldferm$, 
the gauge field strengths $\fldstr$
or their (multiple) covariant derivatives $\cder$
(we hide the spacetime and internal indices).
All fields are evaluated at a common point in spacetime
which we need not specify further for the enumeration.
Due to the trace, the definition of the local operators is invariant 
w.r.t.\ (graded) cyclic shifts $\fld_k\to\fld_{k+1}$,
$\fld_n\to\fld_1$.

In enumerating the local operators one should take 
the (quantum) equations of motions of the fields into account. 
For example, $\cder^2\fldscl$ can be expressed 
through products of the fields such as $\fldscl^3$ or $\fldferm^2$. 
Such combinations are already accounted for in \eqref{eq:locopgen},
so we can discard the term $\cder^2\fldscl$ (irrespectively of
the precise form of the quantum equation of motion).
Similarly, the terms $\cder\cdot\fldferm$ and $\cder\cdot\fldstr$, 
as well as $\cder\wedge\fldstr$ and $\cder\wedge\cder$ can be dropped.
A minimal basis for the fields $\fld$ can be expressed most conveniently 
using spinor indices $\alpha,\beta,\ldots=1,2$ 
and $\dot\alpha,\dot\beta=1,2$ for the Lorentz algebra 
$\alg{so}(3,1)=\alg{sl}(2,\Complex)$ as well as
spinor indices $a,b,\ldots=1,2,3,4$ 
for the the internal algebra $\alg{so}(6)=\alg{su}(4)$.
It turns out that in our basis Lorentz spinor indices are totally symmetric
while internal spinor indices are totally antisymmetric.
Such a basis can be represented through 
states of a supersymmetric oscillator \cite{Gunaydin:1984fk}
with two plus two bosonic operators $\osc{a}^{\dagger \alpha},\osc{b}^{\dagger \dot\alpha}$
and four fermionic operators $\osc{d}^{\dagger a}$. 
Then the various fields of $\superN=4$ SYM decompose as follows
\ifarxiv
\[
\fldstr\sim \osc{b}^\dagger\osc{b}^\dagger,
\quad
\fldferm\sim \osc{b}^\dagger\osc{d}^\dagger,
\quad
\fldscl\sim \osc{d}^\dagger\osc{d}^\dagger,
\quad
\bar\fldferm\sim \osc{a}^\dagger\osc{d}^\dagger\osc{d}^\dagger\osc{d}^\dagger,
\quad
\bar\fldstr\sim \osc{a}^\dagger\osc{a}^\dagger\osc{d}^\dagger\osc{d}^\dagger\osc{d}^\dagger\osc{d}^\dagger,
\quad
\cder\sim \osc{a}^\dagger\osc{b}^\dagger,
\]
\else
\[\begin{array}[b]{rclcrclcrcl}
\fldstr\earel{\sim} \osc{b}^\dagger\osc{b}^\dagger,
&&
\fldferm\earel{\sim} \osc{b}^\dagger\osc{d}^\dagger,
&&
\fldscl\earel{\sim} \osc{d}^\dagger\osc{d}^\dagger,
\\[0.5ex]
\cder\earel{\sim} \osc{a}^\dagger\osc{b}^\dagger,
&&
\bar\fldferm\earel{\sim} \osc{a}^\dagger\osc{d}^\dagger\osc{d}^\dagger\osc{d}^\dagger,
&&
\bar\fldstr\earel{\sim} \osc{a}^\dagger\osc{a}^\dagger\osc{d}^\dagger\osc{d}^\dagger\osc{d}^\dagger\osc{d}^\dagger,
\end{array}\]
\fi
where we have suppressed the indices.
Note that all physical fields in \eqref{eq:locopgen} are uncharged w.r.t.\ the operator
\[
\gen{C}=
2
+N_{\osc{a}}
-N_{\osc{b}}
-N_{\osc{d}},
\]
%
where the $N_{\osc{a},\osc{b},\osc{d}}$ measure the occupation numbers
of the oscillators $\osc{a},\osc{b},\osc{d}$.
For local operators one introduces further indices for the sites, e.g.\
\[
\Tr \fldscl^{ab}\fldscl^{cd}
\sim
\osc{d}^{\dagger  a}_1 \osc{d}^{\dagger  b}_1 
\osc{d}^{\dagger  c}_2 \osc{d}^{\dagger  d}_2 
\state{0}.
\]
Note that on the r.h.s.\ cyclicity is automatic while on the 
l.h.s.\ it must be imposed by hand.
Altogether we have seen that local operators
can be expressed through states of a supersymmetric harmonic oscillator
subject to a charge and a cyclicity constraint.

\subsection{One-Loop Hamiltonian}

\begin{figure}
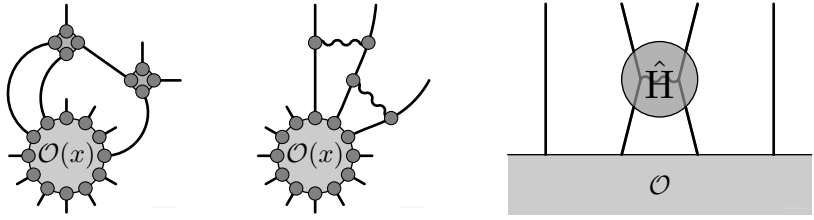
\centering
\includegraphicsbox[scale=\ifarxiv 1.0\else 0.8\fi]{FigChainActNonPlanar.mps}
\qquad
\includegraphicsbox[scale=\ifarxiv 1.0\else 0.8\fi]{FigChainActPlanar.mps}
\qquad
\includegraphicsbox[scale=\ifarxiv 1.0\else 0.8\fi]{FigChainActHam.mps}
\caption{
Non-planar (left) and planar (middle)
gluing of interactions to a local operator $\Op$.
The planar Hamiltonian $\ham$ acts on a pair of nearest neighbours (right)
when zooming into the trace structure of $\Op$.}
\label{fig:ChainActions}
\end{figure}

The anomalous dimensions of local operators
originate from the divergent contributions 
to their two-point functions. 
They are therefore captured by the 
renormalisation matrix $Z$. 
More precisely, the matrix of anomalous dimensions
for the $\Op_A$ 
is given by the logarithmic derivative of $Z$ 
w.r.t.\ the logarithm of the renormalisation scale $\mu$
\[
\delta\gen{D}\sim Z^{-1}\frac{\mu\,d Z}{d\mu}\,.
\]
The eigenvalues of the matrix $\delta\gen{D}$ represent the 
quantum corrections $\delta D_A$ in the scaling dimensions 
$D_A$ of the eigen-operators $\tilde\Op_A$.
The matrix can be interpreted as a Hamiltonian
of a quantum mechanical system:
It acts on the states in a systematic fashion 
determined by connected Feynman diagrams 
attached to the fields constituting the local operators,
see \figref{fig:ChainActions}.
The planar limit suppresses crossing lines in Feynman diagrams, 
therefore $\delta\gen{D}$ acts on a set of adjacent 
fields along the single-trace state \eqref{eq:locopgen}.

The number of fields involved in the action of $\delta\gen{D}$ 
increases with the loop order. 
At the leading one-loop order the action is
between nearest neighbours, cf.\ \figref{fig:ChainActions}
\[
\delta\gen{D}^{(1)}=\ham=\sum_{k=1}^n \ham_{k,k+1},
\]
and it can be interpreted as the Hamiltonian of a quantum spin chain \cite{Minahan:2002ve}.
The one-loop Hamiltonian is invariant under the free 
superconformal symmetries. 
The representation of the latter on the fields 
can be expressed conveniently in the oscillator framework
using bilinears in the operators $(\osc{a}^\dagger,\osc{b},\osc{d})$
and $(\osc{a},\osc{b}^\dagger,\osc{d}^\dagger)$ \cite{Gunaydin:1984fk}
\[\label{eq:RepLocal}
\begin{array}[b]{rclcrcl}
\gen{L}^\alpha{}_\beta\eq \osc{a}^{\dagger\alpha}\, \osc{a}_\beta
-\half\, \delta^\alpha_\beta \osc{a}^{\dagger \gamma}\, \osc{a}_\gamma,
&&
\gen{\bar L}^{\dot \alpha}{}_{\dot \beta}\eq 
\osc{b}^{\dagger \dot \alpha}\,\osc{b}_{\dot \beta}
-\half\,\delta^{\dot \alpha}_{\dot \beta} \osc{b}^{\dagger \dot \gamma}\,\osc{b}_{\dot \gamma},
\\[0.5ex]
\gen{D}\eq \half \osc{a}^{\dagger \gamma}\, \osc{a}_\gamma+\half \osc{b}_{\dot \gamma}\, \osc{b}^{\dagger \dot \gamma},
&&
\gen{R}^a{}_b\eq \osc{d}^{\dagger a}\osc{d}_b-\quarter\,\delta^a_b\osc{d}^{\dagger c}\,\osc{d}_c,
\\[0.5ex]
\bar{\gen{Q}}^{\dot \alpha}_{b}\eq \osc{b}^{\dagger \dot \alpha}\, \osc{d}_b,
&&
\bar{\gen{S}}_{\dot \alpha}^b\eq \osc{d}^{\dagger b}\, \osc{b}_{\dot \alpha} ,
\\[0.5ex]
\gen{Q}^{\beta a}\eq \osc{a}^{\dagger \beta}\, \osc{d}^{\dagger a},
&&
\gen{S}_{\beta a}\eq \osc{a}_\beta\, \osc{d}_a ,
\\[0.5ex]
\gen{P}^{\beta\dot\alpha}\eq \osc{a}^{\dagger \beta}\, \osc{b}^{\dagger \dot\alpha} ,
&&
\gen{K}_{\beta\dot\alpha}\eq \osc{a}_\beta\, \osc{b}_{\dot\alpha}.
\end{array}
\]
Single-trace states \eqref{eq:locopgen} transform in tensor product
representations of the above. 

Invariance under free superconformal symmetry 
imposes strong constraints on $\ham$. 
The crucial observation is that the tensor product of
two field representations decomposes into a 
sequence of irreducible representations 
distinguished by their overall superconformal spin $j$.
The latter can be measured 
using the quadratic Casimir of $\alg{psu}(2,2|4)$ 
in analogy to the total spin of $\alg{su}(2)$.
Symmetry demands that that the Hamiltonian 
has a common eigenvalue for all components 
of an irreducible multiplet.%
\footnote{This holds for multiplets of multiplicity 1; 
for higher multiplicity $n$, 
invariance allows an action equivalent to a $n\times n$ matrix.}
Hence it suffices to specify the eigenvalues,
and we can write the nearest-neighbour Hamiltonian as
\cite{Beisert:2003jj}
\[\label{eq:HamProj}
\ham_{k,k+1}=\sum_{j=0}^\infty c_j \proj_{k,k+1;j}.
\]
Here the operator $\proj_{k,k+1;j}$ projects a two-particle state
to its components with superconformal spin $j$. 
Now there are several ways to determine the unspecified eigenvalues $c_j$:
Direct calculation in the one-loop quantum field theory 
shows that the coefficients are given by the
elements of the harmonic series
\cite{Beisert:2003jj,Kotikov:2000pm,Kotikov:2001sc,Dolan:2001tt}
\[\label{eq:HamCoeff}
c_j\sim h(j)=\sum_{k=1}^j\frac{1}{k}=\psifn(j+1)-\psifn(1),
\qquad
\psifn(z)=\frac{\gammafn'(z)}{\gammafn(z)}\,.
\]
Two other methods of determining the coefficients purely algebraically
are described in the following two subsections.

The analog of the above Hamiltonian for quasi-partonic operators in QCD 
is very similar, and it has a particular feature 
which was noticed in \cite{Braun:1998id,Braun:1999te,Belitsky:1999ru},
see \cite{Belitsky:2004cz} for a review. 
Namely, the appearance of the digamma function $\psifn$
hints at \emph{integrability}, cf.\ \cite{Faddeev:1996iy}.
In particular, Lipatov realised in \cite{Lipatov:1997vu,Lipatov:1998as}
that for $\superN=4$ SYM the Hamiltonian is particularly simple, 
and also made the prophetic connection 
to the newly proposed AdS/CFT correspondence to strings on $AdS_5\times S^5$.
Several years later and in a different context, 
integrability of the one-loop Hamiltonian 
was rediscovered in \cite{Minahan:2002ve,Beisert:2003yb}.
Most importantly, it was also put to use by establishing a set of Bethe equations 
to determine the spectrum of planar one-loop anomalous dimensions very efficiently. 
In particular, the thermodynamic limit of long chains, $n\to\infty$, 
became accessible \cite{Minahan:2002ve,Beisert:2003xu,Beisert:2003ea} 
and could be compared to results from string theory
\cite{Berenstein:2002jq,Gubser:2002tv,Frolov:2002av},%
\footnote{It turned out only later that the matching was more
of a coincidence than a confirmation for AdS/CFT
due to an order of limits issue, see \protect\cite{Beisert:2005cw}.}
see \cite{Beisert:2004ry,Plefka:2005bk,Arutyunov:2009ga} for reviews.

\subsection{Leading-Order Yangian}

Integrable spin chains with manifest Lie algebra symmetry $\alg{g}$
typically have a Yangian algebra $\yangian$ underlying their structure
\cite{Drinfeld:1985rx}. 
The Yangian is a quantum algebra based on (half of) the affine extension 
of the Lie algebra. 
That is to say, next to the Lie generators $\gen{J}^A$,
there are level-one Yangian generators $\genY{J}^A$. 
These obey similar commutation relations as the Lie generators, namely%
\footnote{For reasons of clarity we treat all generators to be bosonic,
the generalisation to superalgebras by insertion of appropriate sign factors is straight-forward.}
\[\label{eq:YangAdj}
\comm{\gen{J}^A}{\gen{J}^B}=F^{AB}_C \gen{J}^C,
\qquad
\comm{\gen{J}^A}{\genY{J}^B}=F^{AB}_C \genY{J}^C,
\]
from which two sets of Jacobi-identities follow. 
However, a third Jacobi-identity involving two Yangian generators
is quantum-deformed to the following Serre relation 
\[\label{eq:YangSerre}
\ifarxiv\else\begin{array}[b]{l}\fi
\bigcomm{\comm{\gen{J}^A}{\genY{J}^B}}{\genY{J}^C}+
\bigcomm{\comm{\gen{J}^B}{\genY{J}^C}}{\genY{J}^A}+
\bigcomm{\comm{\gen{J}^C}{\genY{J}^A}}{\genY{J}^B}
\ifarxiv\else\\[0.5ex]\qquad\qquad\qquad\qquad\mathord{}\fi
=F^{AG}_D F^{BH}_E F^{CI}_F F_{GHI} \gen{J}^{\{D}\gen{J}^E\gen{J}^{F\}}.
\ifarxiv\else\end{array}\fi
\]

A representation of a Lie algebra can sometimes be lifted to 
an evaluation representation of the corresponding Yangian. 
For these, $\gen{J}$ acts as in the Lie algebra and 
$\genY{J}\simeq u\gen{J}$ with $u$ the spectral parameter
of the evaluation representation.
Clearly, the two commutation relations \eqref{eq:YangAdj} are satisfied
automatically, but in addition the r.h.s.\ 
of the the Serre relation \eqref{eq:YangSerre} must vanish.
This is true for the above superconformal representation \cite{Dolan:2004ps}, 
consequently the spin chain transforms in a representation of the Yangian.
Due to homogeneity of the spin chain, 
the spectral parameters of all sites should be equal.

\begin{figure}
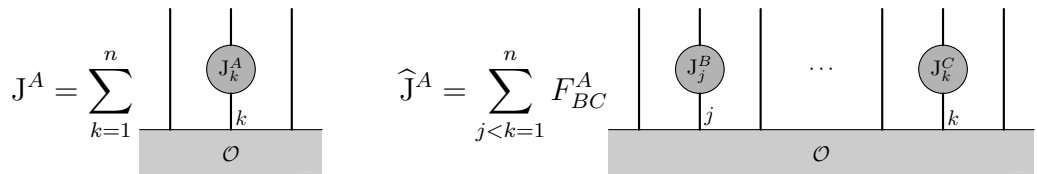
\centering
$\displaystyle\gen{J}^A=
\sum_{k=1}^n\includegraphicsbox[scale=\ifarxiv 0.8\else 0.6\fi]{FigChainFreeConf.mps}$
\qquad
$\displaystyle\genY{J}^A=\sum_{j<k=1}^n F^A_{BC}\includegraphicsbox[scale=\ifarxiv 0.8\else 0.6\fi]{FigChainFreeYang.mps}$
\caption{Action of the free superconformal $\gen{J}^A$ and Yangian 
$\genY{J}^A$ generators on spin chain state alias a local operators $\Op$.}
\label{fig:ChainFreeAction}
\end{figure}

In addition to multiplication, a quantum algebra has a comultiplication operation 
$\copro:\yangian\to\yangian\otimes\yangian$ with
\[\label{eq:Copro}
\copro(\gen{J}^A)=\gen{J}^A\otimes 1+1\otimes\gen{J}^A,
\qquad
\copro(\genY{J}^A)=
\genY{J}^A\otimes 1+1\otimes\genY{J}^A
+
F^A_{BC} \gen{J}^B\otimes \gen{J}^C.
\]
It is compatible with the multiplication,
in particular with the Serre relation \eqref{eq:YangSerre}.
Its main purpose is to define tensor products of representations,
i.e.\ it determines how the algebra acts on the spin chain.
The action on the tensor product of $n$ fields is determined by $\copro^{n-1}(\gen{J})$
\[\label{eq:MultiCopro}
\copro^{n-1}(\gen{J}^A)=\sum_{k=1}^n \gen{J}^A_k,
\qquad
\copro^{n-1}(\genY{J}^A)=
\sum_{k=1}^n \genY{J}^A_k
+
F^A_{BC}\sum_{j<k=1}^n \gen{J}^B_j\gen{J}^C_k.
\]
When using an evaluation representation with homogeneous evaluation 
parameter $u$, we see that the first term in the action of $\genY{J}^A$
equals the superconformal action $u\gen{J}^A$;
therefore nothing is lost by fixing $u$ to a particular value, e.g.\ $u=0$.
Note that while the representation of the Lie generators $\gen{J}$ follows
the usual pattern for tensor products,
the representation of Yangian generators $\genY{J}$ non-trivially 
combines the various sites of the chain.
The action of $\gen{J}^A$ and $\genY{J}^A$ is depicted in \figref{fig:ChainFreeAction}.

An integrable Hamiltonian $\ham$ is invariant under the Lie symmetries $\gen{J}$, 
but it is typically not exactly invariant under the Yangian generators $\genY{J}$.
It commutes up to a difference of two terms \cite{Dolan:2003uh}
\[\label{eq:YangHam}
\comm{\copro(\gen{J}^A)}{\ham_{12}}=0,\qquad
\comm{\copro(\genY{J}^A)}{\ham_{12}}\sim\gen{J}^A_1-\gen{J}^A_2,
\]
On a chain with $n$ sites the commutator yields only boundary terms $\gen{J}_1-\gen{J}_n$
essentially because periodic boundary conditions are not compatible with the
definition of the Yangian. This means that the spectrum of $\ham$ does
not organise into multiplets of the Yangian, but merely of the Lie algebra. 
Nevertheless one can consider the Yangian to be a symmetry of the 
(bulk) Hamiltonian, because commutation (up to boundary terms) 
does yield non-trivial constraints on $\ham$ which guarantee its integrability.
In particular, commutation requires the following recursion relation 
for the coefficients $c_j$ of the Hamiltonian \cite{Dolan:2003uh}
\[
c_{j}=c_{j-1}+\frac{1}{j}\,.
\]
This relation is precisely satisfied by the coefficients from 
field theory \eqref{eq:HamCoeff}, 
and hence planar one-loop $\superN=4$ SYM is integrable.

\subsection{Higher Loops}

Going to higher loops the above picture changes although integrability
apparently remains valid.
The symmetry generators as well as the Hamiltonian receive corrections
in the coupling constant
\[
\gen{J}(g)=\sum_{k=0}^\infty g^k\gen{J}^{(k/2)},\ifarxiv\qquad\else\quad\fi
\genY{J}(g)=\sum_{k=0}^\infty g^k\genY{J}^{(k/2)},\ifarxiv\qquad\else\quad\fi
\ham(g)=\sum_{k=0}^\infty g^k\ham^{(k/2)}.
\]
The structure of the operators must remain compatible with planar
Feynman diagrams, therefore an operator at $\order{g^k}$ involves at
most $k+2$ ingoing plus outgoing fields, see \figref{fig:LongRange}.
In particular, the number of sites of the chain is allowed to fluctuate.
\begin{figure}\centering
$\gen{J}(g)=
    \includegraphicsbox[scale=\ifarxiv 0.6\else 0.4\fi]{FigChainJ0.mps}
+  g\includegraphicsbox[scale=\ifarxiv 0.6\else 0.4\fi]{FigChainJ1a.mps}
+  g\includegraphicsbox[scale=\ifarxiv 0.6\else 0.4\fi]{FigChainJ1b.mps}
+g^2\includegraphicsbox[scale=\ifarxiv 0.6\else 0.4\fi]{FigChainJ2b.mps}
+g^2\includegraphicsbox[scale=\ifarxiv 0.6\else 0.4\fi]{FigChainJ2c.mps}
+\ldots
$
\caption{Perturbative action of a superconformal generator
on a spin chain state.
The deformations involve long-range and dynamic interactions.}
\label{fig:LongRange}
\end{figure}

Despite the above deformations of the representations, 
the algebra relations \eqref{eq:YangAdj} and \eqref{eq:YangSerre}
should remain unchanged. Generally this involves cancellations
between products of terms at various orders. 
These cancellations leave some space for ambiguities, 
and unfortunately the deformations cannot be defined uniquely. 
It turns out that the ambiguities correspond to perturbative
similarity transformations $\gen{J}\to \mathcal{X}\gen{J}\mathcal{X}^{-1}$ of
the generators which leave all algebra relations invariant.%
\footnote{Ambiguities (e.g.\ of ordering) 
are a generic problem of quantum algebras,
which is also the reason why the Serre relation 
\protect\eqref{eq:YangSerre} is 
not formulated in the form of
$\comm{\genY{J}^A}{\genY{J}^B}=\ldots$
analogously to \protect\eqref{eq:YangAdj}.
Hence quantum algebras are typically defined modulo certain types of deformations.}
Only at low orders the set of permissible similarity transformations
is empty and the algebra becomes unique. 

\begin{figure}
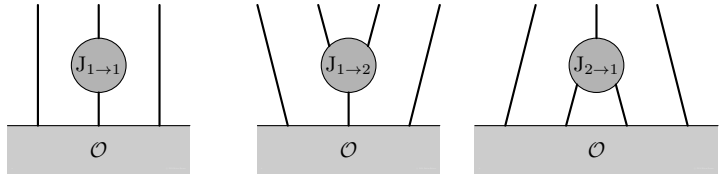
\centering
\includegraphicsbox[scale=\ifarxiv 0.8\else 0.6\fi]{FigChainJ0.mps}\qquad
\includegraphicsbox[scale=\ifarxiv 0.8\else 0.6\fi]{FigChainJ1a.mps}\quad
\includegraphicsbox[scale=\ifarxiv 0.8\else 0.6\fi]{FigChainJ1b.mps}
\caption{Free action of a generic superconformal generator $\gen{J}$
(left)
and leading-order corrections
to $\gen{Q},\bar{\gen{Q}},\gen{P}$ (middle) and $\gen{S},\bar{\gen{S}},\gen{K}$ (right).}
\label{fig:1to2}
\end{figure}

We have seen that symmetry determines the one-loop Hamiltonian \eqref{eq:HamProj} 
up to a sequence of coefficients $c_j$. 
It turns out that the higher-loop corrections impose even stronger constraints: 
The point is that the Hilbert space of the spin chain decomposes 
into irreducible multiplets which are distinguished by their scaling dimension
(among other quantum numbers).
For the free superconformal algebra, 
the multiplets can be of short/atypical or of long/typical type
\cite{Dobrev:1985qv}. 
Short multiplets must have (half) integral superconformal scaling dimension
while long multiplets can have irrational scaling dimensions.
However, the Hamiltonian attributes anomalous dimensions to almost all
irreducible multiplets, long or short. 
Considering the spin chain Hamiltonian as the radiative correction $\delta\gen{D}$
to the dilatation generator $\gen{D}$ seemingly leads to a paradox. 
It is resolved if the right combination of short multiplets 
join to form a long multiplet.%
\footnote{A similar mechanism is required for the Higgs effect 
where a massless vector and a massless scalar combine into a massive vector.}
This can only work if the short multiplets have coincident one-loop
anomalous dimensions, which thus puts constraints on $\ham$. 
On the level of the algebra, the joining of short multiplets
into a long one is achieved through deformations of the 
superconformal generators $\gen{Q},\bar{\gen{Q}},\gen{P}$
and $\gen{S},\bar{\gen{S}},\gen{K}$ at order $\order{g}$.
These map one site to two or vice versa, cf.\ \figref{fig:1to2}.
The algebra turns out to completely determine them. 
Invariance of the Hamiltonian then fixes the coefficients
to the values of field theory \eqref{eq:HamCoeff}
\cite{Beisert:2004ry}
\[
c_j\sim \sum_{k=1}^j \frac{1}{k}\,.
\]
It is curious to see that integrability as well as higher-loop 
consistency lead to precisely the same constraints of the one-loop
Hamiltonian. On the one hand, one can view it as a consistency 
condition of higher-loop integrability, but on the other hand, 
the semantic relation between the two approaches remains somewhat obscure.

\begin{figure}
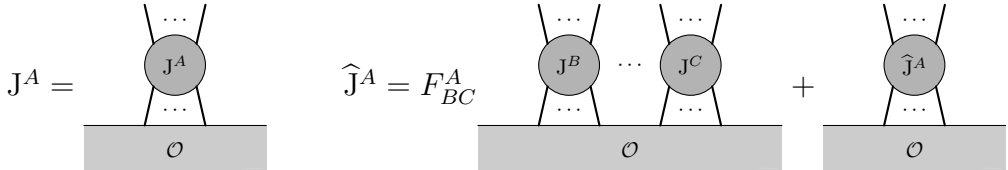
\centering
 $\gen{J}^A=\includegraphicsbox[scale=\ifarxiv 0.8\else 0.6\fi]{FigChainPertConf.mps}$
\qquad
$\genY{J}^A=F^A_{BC}
\includegraphicsbox[scale=\ifarxiv 0.8\else 0.6\fi]{FigChainPertYang1.mps}
+\includegraphicsbox[scale=\ifarxiv 0.8\else 0.6\fi]{FigChainPertYang2.mps}$
\caption{Structure of the deformed representation
of the superconformal and Yangian algebra.
The bilocal contributions to the Yangian are determined by the 
superconformal generators while the local contributions can be viewed
as a short-distance regularisation thereof.}
\label{fig:PertAction}
\end{figure}

The leading-order deformation of the generators was established explicitly 
in \cite{Beisert:2003ys,Zwiebel:2005er}
for closed sectors of the full theory.
In these sectors the construction was also continued by a 
few more perturbative orders. 
Knowledge of the higher-loop Hamiltonian 
revealed first strong hints that integrability
survives \cite{Beisert:2003tq} in perturbation theory.
It was later shown that also the action of the Yangian can be deformed appropriately
\cite{Serban:2004jf,Zwiebel:2006cb,Beisert:2007jv,Bargheer:2008jt,Bargheer:2009xy},
which establishes integrability rigorously at a given perturbative order. 
We note that the structure of the deformed Yangian action 
always follows a pattern analogous to the coproduct rule \eqref{eq:Copro}:
It consists of a bi-local combination of superconformal 
representations (properly expanded at each order of perturbation theory)
and a local contribution which can be viewed
as a short-distance regularisation of the bi-local term, 
see \figref{fig:PertAction}.

\section{Scattering Amplitudes}
\label{sec:scattering}

A different type of observable which plays an important role in 
quantum field theories is the scattering matrix. Integrability
has also been observed for $\superN=4$ SYM in this context, 
and apparently it leads
to substantial simplifications in their structure. We now
review scattering amplitudes and their Yangian symmetry.

\subsection{Framework}

A scattering amplitude of $n$ particles
is a function of the particle momenta $p_k$, 
spins or helicities, flavours 
and gauge degrees of freedom $A_k$.
Statistics requires that this function is (graded) symmetric
under the simultaneous interchange of all quantum numbers
associated to any pair of particles.
This symmetry can be enforced by summing over all (graded) permutations
of particles with the associated quantum numbers
\[
\Amp\supup{full}_{1\ldots n}=
\sum_{\pi\in S_n}
\Amp\supup{ordered}_{\pi(1)\ldots \pi(n)}.
\]

\begin{figure}
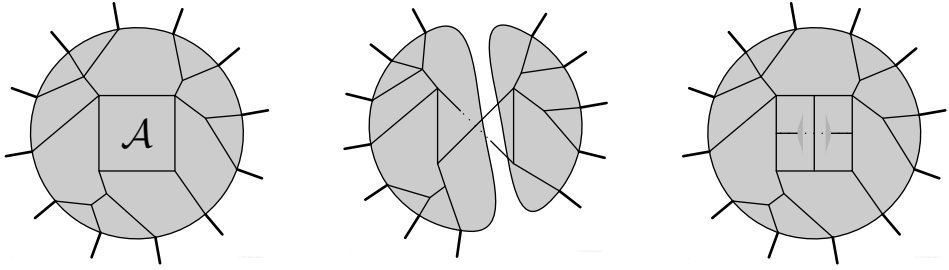
\centering
\includegraphicsbox[scale=\ifarxiv 1.0\else 0.8\fi]{FigAmpTraceSingle.mps}
\qquad
\includegraphicsbox[scale=\ifarxiv 1.0\else 0.8\fi]{FigAmpTraceDouble.mps}
\qquad
\includegraphicsbox[scale=\ifarxiv 1.0\else 0.8\fi]{FigAmpTraceTorus.mps}
\caption{Colour-ordered scattering amplitudes expanded
into $\grp{U}(N\indup{c})$ traces: 
single-trace (left), 
multi-trace (middle) 
and non-planar (right) contributions.}
\label{eq:scatteringtraces}
\end{figure}

In $\superN=4$ SYM all particles transform in the adjoint
representation of the gauge group.
The gauge indices for a $\grp{U}(N\indup{c})$ gauge group
can be expanded in a basis of traces of $\grp{U}(N\indup{c})$ 
generators $T_k:=T^{A_k}$ in the fundamental representation,
see \figref{eq:scatteringtraces}
\<
\Amp\supup{ordered}_{1\ldots n}
\eq
\frac{1}{n}\Tr(T_1\ldots T_n)
\Amp\supup{single-trace}_{1\ldots n}
\nl
+\sum_{k=1}^n
\frac{1}{2k(n-k)}\Tr(T_{1}\ldots T_{k})\,
\Tr(T_{k+1}\ldots T_{n})\,
\Amp\supup{double-trace}_{1\ldots k|k+1\ldots n}
\nl
+\ldots.
\>
The prefactors of $1/k$ and $1/2$ are the appropriate symmetry factors
for $\Integers_k$ cyclic and $S_2$ permutation symmetry.
Note that the various multi-trace or \emph{colour-ordered} contributions to the amplitude
now just depend on the particle momenta, spins/helicities and flavours, but not 
on the gauge structure anymore.

Next, all particles in $\superN=4$ SYM are massless. 
The on-shell momenta $p_k$ are light-like and can be
represented as bilinear combinations of bosonic spinors
$\lambda^\beta$ and $\tilde\lambda^{\dot\alpha}$ 
\cite{Berends:1987me}
\[
 p^{\beta\dot\alpha}
=\sigma^{\beta\dot\alpha}_\mu p^\mu 
=\lambda^\beta \tilde\lambda^{\dot\alpha}.
\]
The two spinors are complex conjugates, 
$\tilde\lambda^{\dot\alpha}=\pm(\lambda^\alpha)^\ast$,
where the sign determines the sign of the particle energy.
Furthermore, all flavours of on-shell particles -- 
scalars $\Phi$, fermions $\Psi$ and gluons $\Gamma$ -- 
can be conveniently combined into a field on superspace 
\cite{Nair:1988bq}
\[
\Omega(\lambda,\tilde\lambda,\bar\eta)=
\Gamma
+\bar\eta^a\Psi_a
+\half\bar\eta^a\bar\eta^b\Phi_{ab}
+\sfrac{1}{6}\varepsilon_{abcd}\bar\eta^a\bar\eta^b\bar\eta^c\bar\Psi^{d}
+\sfrac{1}{24}\varepsilon_{abcd}\bar\eta^a\bar\eta^b\bar\eta^c\bar\eta^d\bar\Gamma.
\]
This is useful because we can now scatter the superfield
$\Omega$
instead of the individual fields: 
The colour-ordered amplitudes then turn into plain function 
on the configuration superspace
parametrised by $\lambda_k$, $\tilde\lambda_k$ and $\bar\eta_k$
or, collectively, $\Lambda_k$ 
\[
\Amp\supup{colour-ordered}(\Lambda_1,\ldots,\Lambda_n).
\]
In particular, the flavour indices have been traded in
completely for expansion coefficients in the $\bar\eta_k$.
Also the helicity is determined by the flavour in $\superN=4$ SYM,
so that no indices remain. 
Such colour-ordered amplitudes on superspace 
will be the standard objects we shall consider.

Note that the multiplication of $\lambda^\alpha$ by a complex
phase does not change the momentum $p^\mu$.
Such a multiplication is equivalent to a 
rotation about the particle momentum,
and thus the amplitude must 
transform according to the particle helicity.
In effect, the amplitude is constrained by 
\[
\Amp(\ldots,\Lambda_k,\ldots)=
e^{2i\varphi}\Amp(\ldots,e^{i\varphi}\Lambda_k,\ldots),
\ifarxiv\qquad\else\quad\fi
e^{i\varphi}(\lambda,\tilde\lambda,\eta)
:=(e^{i\varphi}\lambda,e^{-i\varphi}\tilde\lambda,e^{i\varphi}\eta).
\]
Put differently, the differential operator
\[
\gen{C}_k=
2+\lambda^\alpha_k\frac{\partial}{\partial\lambda^\alpha_k}
-\tilde\lambda^\alpha_k\frac{\partial}{\partial\tilde\lambda^{\dot\alpha}_k}
-\bar\eta^{a}_k\frac{\partial}{\partial\bar\eta^a_k}
\]
acting on any leg $k$ annihilates the amplitude.

Furthermore, one can classify amplitudes by an operator $\gen{B}$
which effectively measures the overall helicity of the particles
\[
\Amp_n=\sum_{k=2}^{n-2}\Amp_{n,k},
\qquad
\gen{B}\Amp_{n,k}=4k\Amp_{n,k},
\qquad
\gen{B}=\sum_{k=1}^n 
\bar\eta^{a}_k\frac{\partial}{\partial\bar\eta^a_k}=N_{\bar\eta}\,.
\]
Note that due to $\alg{su}(4)$-invariance of the amplitude,
the $\bar\eta$'s can only appear in groups of four and 
due to supersymmetry there must be between $8$ and $4(n-2)$ of them.
The amplitude with the minimum number of eight $\bar\eta$'s is called MHV
and it is typically the simplest among those
with the same number of legs.

\subsection{Tree-Level Amplitudes}
\label{sec:trees}

The MHV amplitudes $\Amp\supup{MHV}_n:=\Amp_{n,2}$ have particularly
simple expressions
\cite{Parke:1986gb,Berends:1987me}
\[\label{eq:MHVtree}
\Amp\supup{MHV}_n=\frac{\delta^4(P_n)\,\delta^8(Q_n)}{\sprod{1}{2}\cdots\sprod{n}{1}}\,
\]
with the overall momentum $P_n$ and its fermionic partner $Q_n$
\[
P^{\beta\dot\alpha}_n=\sum_{k=1}^n 
\lambda^{\beta}_k
\tilde\lambda^{\dot\alpha}_k,
\qquad
Q^{\beta a}_n=\sum_{k=1}^n 
\lambda^{\beta}_k
\bar\eta^{a}_k.
\]
Furthermore invariants of the spinors are obtained by contraction
with the antisymmetric invariant tensor $\varepsilon_{\alpha\beta}$ 
or $\varepsilon_{\dot\alpha\dot\beta}$
\[
\sprod{\lambda}{\mu}:=\varepsilon_{\alpha\beta}\lambda^\alpha\mu^{\beta},
\qquad
\cprod{\tilde\lambda}{\tilde\mu}:=\varepsilon_{\dot\alpha\dot\beta}\tilde\lambda^{\dot\alpha}\tilde\mu^{\dot\beta}.
\]
We abbreviate $\sprod{\lambda_j}{\lambda_k}$ as $\sprods{j}{k}$. 
We can use the above expression $\Amp\supup{MHV}$ to confirm 
$\superN=4$ superconformal invariance of scattering amplitudes. 
The generators of $\alg{psu}(2,2|4)$ acting on a single free field 
take particularly simple expressions 
using the spinor helicity superspace variables
$\lambda,\tilde\lambda,\bar\eta$
\cite{Witten:2003nn}
\[\label{eq:SHconformal}
\begin{array}[b]{@{}rclcrcl@{}}
\gen{L}^\alpha{}_\beta\eq \lambda^\alpha\partial_{\beta}-\half\delta^\alpha_\beta \lambda^\gamma\partial_\gamma,
&&
\gen{\bar L}^{\dot \alpha}{}_{\dot \beta}\eq 
\tilde\lambda^{\dot \alpha}\tilde\partial_{\dot \beta}
-\half\delta^{\dot \alpha}_{\dot \beta} \tilde\lambda^{\dot \gamma}\tilde \partial_{\dot \gamma},
\\[0.5ex]
\gen{D}\eq \half\partial_\gamma\lambda^\gamma+\half\tilde\lambda^{\dot \gamma}\tilde\partial_{\dot \gamma},
&&
\gen{R}^a{}_b\eq \bar\eta^a\bar\partial_b-\quarter \delta^a_b\bar\eta^c\bar\partial_c,
\\[0.5ex]
\bar{\gen{Q}}^{\dot \alpha}_{b}\eq \tilde\lambda^{\dot \alpha} \bar\partial_b ,
&&
\bar{\gen{S}}_{\dot \alpha}^b\eq \bar\eta^b \tilde\partial_{\dot \alpha} ,
\\[0.5ex]
\gen{Q}^{\beta a}\eq \lambda^\beta\bar\eta^a ,
&&
\gen{S}_{\beta a}\eq \partial_\beta\bar\partial_a  ,
\\[0.5ex]
\gen{P}^{\beta\dot\alpha}\eq \lambda^\beta\tilde\lambda^{\dot \alpha} ,
&&
\gen{K}_{\beta\dot\alpha}\eq \partial_\beta\tilde\partial_{\dot \alpha}.
\end{array}
\]
The action on the amplitude is given by the standard tensor product rule
as the sum over all fields
\[
\gen{J}^A=\sum_{k=1}^n \gen{J}^A_k.
\]
Invariance under the Lorentz $\gen{L},\bar{\gen{L}}$ 
and internal $\gen{R}$ rotations is manifest because the 
amplitude \eqref{eq:MHVtree} is constructed only from scalar combinations.

The dilatation generator $\gen{D}$ counts the number of 
$\lambda$'s and $\tilde\lambda$'s. For scaling invariance
the overall number must equal $-2n$. Using the degrees of homogeneity 
of the three components \eqref{eq:MHVtree} 
\[\label{eq:MHVweight}
\delta^4(P)\sim\lambda^{-4}\bar\lambda^{-4},\qquad
\delta^8(Q)\sim\bar\eta^8\lambda^{8},\qquad
\frac{1}{\sprods{1}{2}\ldots\sprods{n}{1}}\sim\lambda^{-2n}.
\]
invariance follows straight-forwardly.
Furthermore, it is clear that each $\gen{C}_k$ annihilates
the amplitude as desired because both delta-functions are invariant,
and the denominator contributes $\lambda_k^{-2}$ for each $k$. 

Next, the translations $\gen{P}$ and the supertranslations $\gen{Q}$
annihilate the amplitude due to the two delta-functions 
$\delta^4(P_n)$ and $\delta^8(Q_n)$
because $\gen{P}$ acts through multiplication by the overall momentum $P_n$ 
(analogously for $\gen{Q}$ and $Q_n$).

Invariance under the conjugate supertranslation is less obvious. 
As it contains a derivative w.r.t.\ $\bar\eta$, it acts non-trivially
only on the fermionic delta-function $\delta^8(Q_n)$ 
\[\label{eq:Qbaract}
\gen{\bar Q}_b^{\dot \alpha}\delta^8(Q_n)
=
\sum_{k=1}^n\tilde\lambda^{\dot\alpha}_k \bar\partial_{b,k}\delta^8(Q_n)
=
\sum_{k=1}^n \lambda_k^\gamma \tilde\lambda^{\dot\alpha}_k 
\frac{\partial \delta^8(Q_n)}{\partial Q^{\gamma b}_n}
=
P^{\gamma\dot\alpha}_n
\frac{\partial \delta^8(Q_n)}{\partial Q^{\gamma b}_n}\,.
\]
Due to the presence of the bosonic delta-function $\delta^4(P_n)$ 
the conjugate supermomentum annihilates the amplitude.
The derivation for invariance under the conjugate superboost $\bar{\gen{S}}$ 
is analogous, but there are important subtleties to be discussed in 
\secref{eq:AmpLoop}.

To show invariance under the conformal boost $\gen{K}$ and superboost $\gen{S}$ 
takes the largest number of steps. The two derivations are analogous and we consider
only the superboost $\gen{S}$. It contains a derivative w.r.t.\ $\bar\eta$
which again only acts on the fermionic delta-function.
By a sequence of transformation we can recombine the 
terms into useful combinations
\<
\gen{S}_{\alpha b}\delta^8(Q_n)
\eq
\sum_{k=1}^n 
\partial_{\alpha,k} \lambda_k^\gamma
\frac{\partial\delta^8(Q_n)}{\partial Q^{\gamma b}_n}
=
\lrbrk{
\gen{L}^\gamma{}_\alpha
+\half\delta^\gamma_\alpha
\sum_{k=1}^n 
(\lambda_k^\delta\partial_{\delta,k}+2)
}
\frac{\partial\delta^8(Q_n)}{\partial Q^{\gamma b}_n}
\nln
\eq
\frac{\partial\delta^8(Q_n)}{\partial Q^{\gamma b}_n}\,
\gen{L}^\gamma{}_\alpha
+
\frac{1}{2}\,
\frac{\partial\delta^8(Q_n)}{\partial Q^{\alpha b}_n}
\lrbrk{
\sum_{k=1}^n 
\lambda_k^\delta\partial_{\delta,k} 
-3+7+2n
}.
\nl
\>
The first step consists in rewriting
$\partial_{\alpha} \lambda^\gamma$
as a Lorentz generator $\gen{L}^\gamma{}_\alpha$.
In the next step these generators are commuted past 
the fermionic delta-function. This picks up $-\sfrac{3}{2}$ 
from the Lorentz generator and $\sfrac{7}{2}$ from the 
weight in $\lambda$.
The point is then that the remaining denominator
and bosonic delta-function 
in $\Amp\supup{MHV}_n$ \eqref{eq:MHVtree} are 
Lorentz invariant have overall weight $\lambda^{-4-2n}$
according to \eqref{eq:MHVweight}. Hence the amplitude is annihilated.

\subsection{Leading-Order Yangian}

In addition to the standard superconformal symmetries 
a new type of superconformal symmetry has recently been discovered 
for planar scattering amplitudes in $\mathcal{N}=4$
\cite{Drummond:2006rz,Bern:2006ew,Bern:2007ct,Drummond:2008vq}. 
Tree amplitudes were shown to be covariant with respect to these \emph{dual superconformal}
transformations \cite{Brandhuber:2008pf,Drummond:2008cr},
and also loop amplitudes appear to be substantially constrained.
The dual superconformal algebras overlaps partially
with the conventional one, and therefore the two algebras 
must close onto a bigger one. 
This algebra turns out to be a Yangian \cite{Drummond:2009fd}.

\begin{figure}
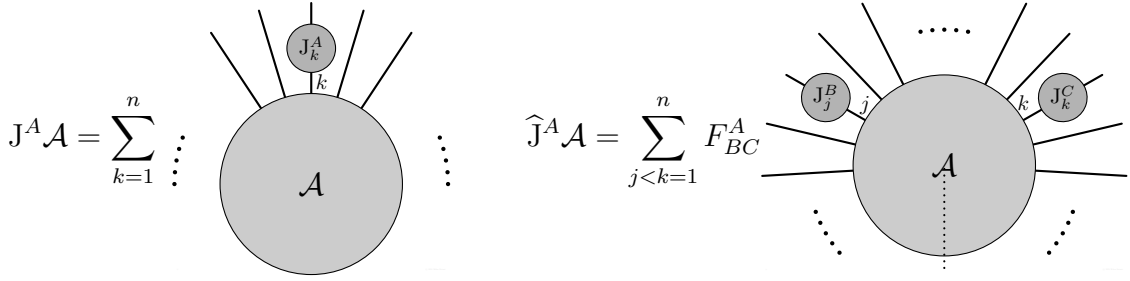
\centering
$\displaystyle\gen{J}^A\Amp=\sum_{k=1}^n\ \includegraphicsbox[scale=\ifarxiv 0.8\else 0.5\fi]{FigAmpConfFree.mps}$
\qquad
$\displaystyle\genY{J}^A\Amp=\sum_{j<k=1}^n F^A_{BC}\includegraphicsbox[scale=\ifarxiv 0.8\else 0.5\fi]{FigAmpYangFree.mps}$
\caption{Action of the free superconformal generator $\gen{J}$
and the Yangian generator $\genY{J}$
on the colour-ordered amplitude $\Amp$.}
\label{fig:treeyang}
\end{figure}

We now wish to extend the superconformal symmetry for amplitudes
to a Yangian algebra. 
The fields transform in the superconformal representation 
specified in \eqref{eq:SHconformal}.
It can be extended to an evaluation representation 
of the Yangian because the Serre relations are satisfied. 
Clearly, all external fields are on equal footing and should 
have coincident evaluation parameter. 
Again, its value does not play an important role because it merely
multiplies the standard conformal generators; we can safely set it to zero.
The representation of the Yangian generators 
from the coproduct \eqref{eq:Copro}
then becomes, 
see also \figref{fig:treeyang},
\[\label{eq:YangActAmp}
\genY{J}^A=\half F^A_{BC} \gen{J}^B \wedge \gen{J}^C,
\qquad
\mbox{where}\qquad
\gen{J}^B\wedge\gen{J}^C:=
\sum_{j<k=1}^n \bigbrk{
\gen{J}^B_j \gen{J}^C_k-\gen{J}^B_k \gen{J}^C_j}.
\]
Invariance of tree amplitudes 
under Yangian symmetry
$\genY{J}^A\Amp=0$
\cite{Drummond:2009fd}
follows from their conventional and dual 
superconformal transformation properties
\cite{Brandhuber:2008pf,Drummond:2008cr}.

Let us demonstrate Yangian invariance of the MHV tree amplitude.
The simplest of the Yangian generators is the level-one
momentum generator $\genY{P}$. 
Due to the adjoint transformation property of the Yangian generators \eqref{eq:YangAdj} 
it suffices to show invariance w.r.t.\ this generator 
in addition to superconformal invariance in order to prove complete Yangian invariance.
The generator takes the explicit form 
\[\label{eq:YangMom}
\genY{P}^{\beta\dot\alpha}=
\gen{P}^{\beta\dot\alpha}\wedge\gen{D}
+\gen{P}^{\delta\dot\alpha}\wedge\gen{L}^\beta{}_\delta
+\gen{P}^{\beta\dot\gamma}\wedge\gen{\bar{L}}^{\dot\alpha}{}_{\dot\gamma}
+\gen{Q}^{\beta c}\wedge\gen{\bar{Q}}^{\dot\alpha}{}_{c}.
\]
This generator has one derivative which acts 
on the amplitude function \eqref{eq:MHVtree}.
The action on delta-functions cancels straight-forwardly 
between the various contributions in \eqref{eq:YangMom}.
What remains is the action on the denominator terms
\<
\genY{P}^{\beta\dot\alpha}\Amp\supup{MHV}_n
\eq
\sum_{j<k}
\lrbrk{
-\lambda^\beta_k\tilde\lambda_j^{\dot\alpha} \frac{\sprod{j}{k+1}}{\sprod{k}{k+1}}
-\lambda^\beta_k\tilde\lambda_j^{\dot\alpha} \frac{\sprod{j}{k-1}}{\sprod{k}{k-1}}
+\lambda^\beta_j\tilde\lambda^{\dot\alpha}_j 
}\Amp\supup{MHV}_n
\nl
+\sum_{j<k}
\lrbrk{
+\lambda^\beta_j\tilde\lambda_k^{\dot\alpha} \frac{\sprod{k}{j+1}}{\sprod{j}{j+1}}
+\lambda^\beta_j\tilde\lambda_k^{\dot\alpha} \frac{\sprod{k}{j-1}}{\sprod{j}{j-1}}
-\lambda^\beta_k\tilde\lambda^{\dot\alpha}_k
}\Amp\supup{MHV}_n.
\ifarxiv\else\nl\fi
\>
Shifting the summation variables appropriately we 
can use the spinor identity
\[
\lambda^\beta_{k}
\sprod{j}{k+1}
-\lambda^\beta_{k+1}
\sprod{j}{k}
=
\lambda^\beta_j
\sprod{k}{k+1}
\]
to combine several terms. What remains turns out to be proportional
to the overall momentum and thus vanishes proving Yangian invariance
for tree MHV amplitudes
\[
\genY{P}^{\beta\dot\alpha}\Amp\supup{MHV}_n
=
\frac{\lambda^\delta_1\lambda^\beta_n+\lambda^\beta_1\lambda^\delta_n}{\sprod{n}{1}}\,
\varepsilon_{\delta\gamma} 
P^{\gamma\dot\alpha}_n
\Amp\supup{MHV}_n=0.
\]

Noting that colour-ordered amplitudes are cyclic,
an important additional consideration is the cyclic behaviour of the Yangian
\cite{Drummond:2009fd}.
The point is that the Yangian generators are typically not invariant
under cyclic shifts: Let us compare the action on sites $1$ through $n$ 
with the action on sites $2$ through $n+1$
\[
\genY{J}^A_{1,n}=\half F^A_{BC}\sum_{k=1}^n\sum_{j=1}^{k-1}\gen{J}_j^B\gen{J}_k^C,
\qquad
\genY{J}^{A}_{2,n+1}=\half F^A_{BC}\sum_{k=2}^{n+1}\sum_{j=2}^{k-1}\gen{J}_j^B\gen{J}_k^C.
\]
These two expressions are not equal, they differ by
\[\label{eq:YangCycl}
\genY{J}^{A}_{2,n+1}-\genY{J}^A_{1,n}
=
-\half F^A_{BC}\acomm{\gen{J}_1^B}{\gen{J}^C}
=
\half F^A_{BC}F^{BC}_D\gen{J}_1^D
-F^A_{BC}\gen{J}_1^B\gen{J}^C.
\]
Hence the action typically maps 
cyclic states to non-cyclic ones. 
More importantly, the action of the Yangian on periodic states
is not uniquely defined; 
it depends on the point where the periodic chain is cut open.

For amplitudes however the situation is better because 
both operators on the r.h.s.\ are symmetries.
The second term vanishes because the amplitude 
is invariant under conventional superconformal symmetry. 
The first term contains $F^A_{BC}F^{BC}_D$ which is proportional
to the dual Coxeter number which equals zero for $\alg{psu}(2,2|4)$.

\subsection{Collinearities and Higher Loops}
\label{eq:AmpLoop}

The discussion of the free superconformal symmetries
in \secref{sec:trees} was not entirely honest
to the end that the amplitude is \emph{not exactly}
invariant under them:
The special superconformal symmetries $\gen{S},\gen{\bar S},\gen{K}$ 
acting on a colour-ordered amplitude
leave behind a distributional remainder supported on 
configurations with a pair of adjacent particles being 
exactly collinear, $p_k\sim p_{k+1}$.
In other words, generic amplitudes are indeed annihilated 
by the free symmetries as discussed above, 
but there exist special configuration where this is not case.
The extra contributions originate in the analog of
\eqref{eq:Qbaract} for $\gen{\bar S}$
from the holomorphic anomaly
in the complex spinor helicity space
\cite{Cachazo:2004by,Cachazo:2004dr}
\[\label{eq:collanom}
\frac{\partial}{\partial\tilde\lambda^{\dot\alpha}}
\,\frac{1}{\sprod{\lambda}{\mu}}
=2\pi\varepsilon_{\dot\alpha\dot\gamma}
\tilde\mu^{\dot\gamma}\sign\bigbrk{E(\lambda)E(\mu)}
\delta^2\bigbrk{\sprod{\lambda}{\mu}}.
\]
The delta-function is supported on collinear spinors
$\lambda,\mu$ or, equivalently, when the associated momenta are collinear.
Luckily these contributions can be compensated
by deforming the representation of $\gen{S},\gen{\bar S},\gen{K}$
\cite{Bargheer:2009qu}.
The additional contributions map one leg of the amplitude
to two or three collinear particles, cf.\ \figref{fig:DynamicInvarianceTree}.
When acting with such an operator on an amplitude
with fewer legs, one can cancel the contributions
from the collinear anomaly. 
Altogether invariance at tree level is recovered 
only when acting on the superposition 
of all amplitudes with arbitrary numbers of legs
(henceforth called \emph{the} amplitude).

\begin{figure}\centering
$\gen{J}\Amp=
\includegraphicsbox[scale=\ifarxiv 0.7\else 0.5\fi]{FigAmpInvTree1.mps}
+\includegraphicsbox[scale=\ifarxiv 0.7\else 0.5\fi]{FigAmpInvTree2.mps}
+\includegraphicsbox[scale=\ifarxiv 0.7\else 0.5\fi]{FigAmpInvTree3.mps}
=0$
\caption{Exact invariance of amplitudes 
under the deformed superconformal representation
at tree level.}
\label{fig:DynamicInvarianceTree}
\end{figure}

Non-invariance under the free superconformal generators turns out to be beneficial 
in several respects. 
While the free superconformal and Yangian generators
only relate amplitudes with a common number of legs, 
the deformations introduce relations between 
amplitudes with different numbers of legs. 
Here the free and deformed generators serve two different purposes:
The free superconformal generators are sensitive to the 
pole-like collinear singularities in the denominator of \eqref{eq:MHVtree} 
through the collinear anomaly \eqref{eq:collanom}. 
The deformed generator provides the residues of the collinear singularities. 
Apart from those inherited from fewer-leg amplitudes, 
further collinear singularities are consequently prohibited by superconformal symmetry. 
In conclusion, conformal symmetry constrains and determines
both the analytical structure and the structure of singularities 
of the amplitude. Together with Yangian symmetry it appears that 
the tree amplitude may be completely determined through symmetry arguments alone!
Although there exist many ways to construct tree amplitudes conveniently,
unique determination by symmetry is an interesting prospect
because uniqueness is automatically inherited to all
perturbative orders \cite{Bargheer:2009qu}.
Nevertheless one has to bear in mind that this also requires understanding how 
the representation is deformed at loop level. 

Concerning the latter issues, 
it appears that the deformations
at loop level depend to a large extent 
on the deformation at tree level 
or a suitable iteration thereof \cite{Sever:2009aa,Beisert:2010gn}. 
In addition there are deformations due to infra-red divergences 
at loop level affecting all of the non-manifest symmetries.
The divergent contributions to the one-loop planar amplitude $\Amp_{n}^{(1)}$
are determined by the tree amplitude $\Amp_{n}^{(0)}$
\[
\Amp_{n}^{(1)}=\hat{Z}^{(1)}A_{n}^{(0)}+\tilde \Amp_{n}^{(1)}
\quad \text{with} \quad 
\hat{Z}^{(1)}=-\sum_{j=1}^{n}\frac{c_{\epsilon}}{\epsilon^{2}}
\left(\frac{s_{j,j+1}}{-\mu^{2}}\right)^{-\epsilon} \, ,
\]
where $s_{j,k}=(p_j+p_k)^2$ and where $\tilde \Amp_{n}^{(1)}$ is finite.
The anomaly of the dilatation operator
due to the presence of the scale $\mu$ is the clearest.
It can be absorbed by a simple one-loop deformation $\gen{D}^{(1)}$
to the free dilatation generator $\gen{D}^{(0)}$
\cite{Beisert:2010gn}
\[
\gen{D}^{(1)}=-\bigcomm{\gen{D}^{(0)}}{\hat{Z}^{(1)}}
=-2\sum_{j=1}^{n}\frac{c_{\epsilon}}{\epsilon}
\left(\frac{s_{j,j+1}}{-\mu^{2}}\right)^{-\epsilon} \, ,
\]
so that $\gen{D}^{(0)}\Amp^{(1)}_n+\gen{D}^{(1)}\Amp^{(0)}_n=0$
because the finite contribution $\tilde \Amp_{n}^{(1)}$ is scale invariant.

\begin{figure}
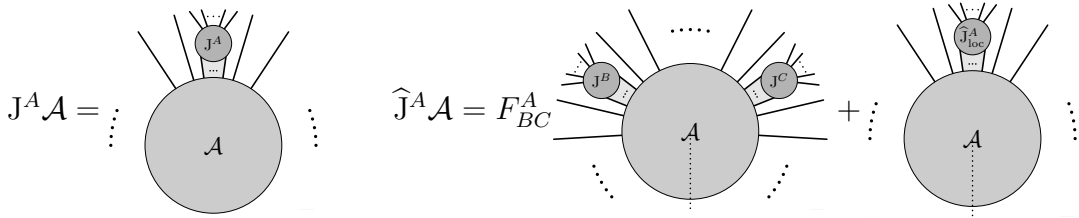
\centering
$\gen{J}^A\Amp=\includegraphicsbox[scale=\ifarxiv 0.6\else 0.45\fi]{FigAmpConfLoc.mps}$
\qquad
$\genY{J}^A\Amp=F^A_{BC}\includegraphicsbox[scale=\ifarxiv 0.6\else 0.45\fi]{FigAmpYangBi.mps}
+\includegraphicsbox[scale=\ifarxiv 0.6\else 0.45\fi]{FigAmpYangLoc.mps}$
\caption{Perturbative action of the superconformal generator $\gen{J}$
and the Yangian generator $\genY{J}$
on the colour-ordered amplitude $\Amp$.}
\label{fig:loopyang}
\end{figure}

The situation for the Yangian momentum generator $\genY{P}$ is similar. 
The anomaly due to the IR divergencies can be absorbed 
by a deformation analogous to the dilatation generator,
but here the finite remainder is anomalous as well
\cite{Drummond:2008vq,Drummond:2008bq,Brandhuber:2008pf,Brandhuber:2009xz,Elvang:2009ya,Brandhuber:2009kh}
\[
(\genY{P}^{(0)})^{\beta\dot\alpha}\, \tilde{\Amp}^{(1)}_n
=2\, \sum_{j=1}^n p_j^{\beta\dot\alpha}\, \log \left(\frac{s_{j,j+1}}{s_{j-1,j}} \right)\Amp^{(0)}_n
.
\]
Note that this anomaly depends only on neighbouring legs. Hence
the total deformation of the Yangian momentum generator reads
\ifarxiv
\[
\genY{P}^{(1)}= 
-\bigcomm{\genY{P}^{(0)}}{\hat{Z}^{(1)}}
+\genY{P}^{(1)}\indup{loc}
\qquad\mbox{with}\qquad
(\genY{P}^{(1)}\indup{loc})^{\beta\dot\alpha} =
2\sum_{j=1}^n(p_j^{\beta\dot\alpha}-p_{j+1}^{\beta\dot\alpha})\,  
\frac{c_{\epsilon}}{\epsilon}
\left(\frac{s_{j,j+1}}{-\mu^{2}}\right)^{-\epsilon} 
.
\]
\else
\<
\genY{P}^{(1)}\eq
-\bigcomm{\genY{P}^{(0)}}{\hat{Z}^{(1)}}
+\genY{P}^{(1)}\indup{loc}
\qquad\mbox{with}\qquad
\nln
(\genY{P}^{(1)}\indup{loc})^{\beta\dot\alpha} \eq
2\sum_{j=1}^n(p_j^{\beta\dot\alpha}-p_{j+1}^{\beta\dot\alpha})\,  
\frac{c_{\epsilon}}{\epsilon}
\left(\frac{s_{j,j+1}}{-\mu^{2}}\right)^{-\epsilon} 
.
\>
\fi
It follows the general structure of perturbative Yangian generators:
The commutator generates the bi-local combinations 
of the deformed generators. 
In this case the dilatation generator $\gen{D}$ is the only
deformed generator among the ones contributing to $\genY{P}$ in \eqref{eq:YangMom}
because the super-Poincar\'e generators $\gen{P},\gen{Q},\gen{\bar Q}$ 
are manifest symmetries. In particular the deformation reads simply, 
see \figref{fig:loopyang}
\[
\genY{P}^{(1)}=\gen{P}^{(0)}\wedge\gen{D}^{(1)}
+\genY{P}^{(1)}\indup{loc}.
\]
The local contribution can be attributed to the anomaly of the finite remainder. 

At higher loops we expect this general picture in \figref{fig:loopyang} 
to remain valid.
There are reasons to believe that the deformation of the 
dilatation $\gen{D}$ and Yangian momentum $\genY{P}$ generator 
at higher loops remains reasonably simple.
For the other superconformal and Yangian generators, however,
the deformation is already substantially more involved even at
one loop \cite{Beisert:2010gn}.
And even though the deformations are known,
it remains to be understood how the superconformal and Yangian 
algebra closes precisely.

\section{Comparison and Summary}
\label{sec:comparison}

The attentive reader will have noticed
that the discussions in \secref{sec:spinchains}
and in \secref{sec:scattering}
were analogous to a large extent.

\subsection{Analogies}

Let us first concentrate on the representation of superconformal symmetry.
The free representations on fields \eqref{eq:RepLocal} and
on external particles \eqref{eq:SHconformal} are equivalent
if one identifies oscillators with spinor-helicity variables
\[
\osc{a}\sim \lambda,\quad
\osc{b}\sim \tilde\lambda,\quad
\osc{d}\sim \bar\eta,\qquad
\osc{a}^\dagger\sim\frac{\partial}{\partial\lambda}\,,\quad
\osc{b}^\dagger\sim\frac{\partial}{\partial\tilde\lambda}\,,\quad
\osc{d}^\dagger\sim\frac{\partial}{\partial\bar\eta}\,.
\]
It is clear that their algebras coincide
and hence the derived representations are equivalent.

This is not surprising because both describe 
free on-shell fields of $\superN=4$ SYM:
The spinor helicity superspace is designed to describe
a field excitation with definite on-shell momentum.
Conversely, 
a finite excitation of the supersymmetric oscillator
describes a component of the fields expanded around
a specific point in spacetime. 
Here, the free equations of motion are imposed through
vanishing of certain components. 
A Fourier transformation translates between the two pictures
\ifarxiv
\[
\fld(x)
\sim\int d^{4}\lambda\,
e^{ix\cdot p(\lambda)}\Omega(\lambda),
\qquad
\Omega(\lambda)
\sim\int d^3x\,
e^{-ix\cdot p(\lambda)}
\bigbrk{
\fld(x,0)
-iE(\lambda)^{-1}\dot\fld(x,0)
}
.
\]
\else
\<
\fld(x)
\earel{\sim}\int d^{4}\lambda\,
e^{ix\cdot p(\lambda)}\Omega(\lambda),
\nln
\Omega(\lambda)
\earel{\sim}\int d^3x\,
e^{-ix\cdot p(\lambda)}
\bigbrk{
\fld(x,0)
-iE(\lambda)^{-1}\dot\fld(x,0)
}
.
\>
\fi
For the reverse transformation it suffices to use a time slice 
of $\fld$ at $t=0$.
It is however clear that for full equivalence between the two pictures 
one has to rely on distributions. 
Consequently the two representations are only equivalent in 
a physicist's sense or under additional assumptions.%
\footnote{It might be interesting to Fourier transform some
of the structures from one picture between position and 
momentum space.}

Once the equivalence of the free superconformal representations is established, 
it becomes straight-forward to lift the equivalence to the Yangian algebra. 
Indeed, the infinite-dimensional algebra 
appears to determine uniquely relevant structures such
as the spin chain Hamiltonian as well as the scattering amplitude.
There is however one crucial difference between the application 
of Yangians to local operators vs.\ scattering amplitudes.
For the former it merely acts as a useful algebraic structure,
cf.\ \eqref{eq:YangHam}, while for the latter it is a true symmetry.
This point will be discussed in more detail in the next subsection.
A related issue is that Yangian symmetry is typically 
incompatible with cyclic symmetry. 
Only for the scattering amplitude 
it respects it due to superconformal invariance
and due to vanishing dual Coxeter number, cf.\ \eqref{eq:YangCycl}. 

We have furthermore seen that the structure of the perturbative 
superconformal and Yangian representation is analogous
in both pictures, cf.\ the pairs
\eqref{eq:MultiCopro,eq:YangActAmp},
\figref{fig:ChainFreeAction,fig:treeyang},
\figref{fig:1to2,fig:DynamicInvarianceTree} and
\figref{fig:PertAction,fig:loopyang}. 
The superconformal deformations act
on several adjacent fields or legs in the
trace of the local operator or colour-ordered amplitude. 
The deformed Yangian representation is constructed as a bi-local 
combination of deformed superconformal representations
plus local terms which can be understood as a short-distance (along the trace) 
regularisation of the bi-local terms.
One notable difference concerns the manifest symmetries
which are not deformed by radiative corrections.
For local operators only the Lorentz and internal symmetries $\gen{L},\gen{\bar L},\gen{R}$ are manifest, 
while for the scattering amplitude the full super-Poincar\'e algebra
including the (super) momentum generators $\gen{Q},\gen{\bar Q},\gen{P}$
are undeformed.%
\footnote{In fact, trying to construct a representation for local operators 
with manifest super-Poincar\'e symmetry
implies vanishing anomalous dimensions \protect\cite{Beisert:2003ys}.}

\subsection{Large-\texorpdfstring{$N\indup{c}$}{Nc} Topology}

\begin{figure}
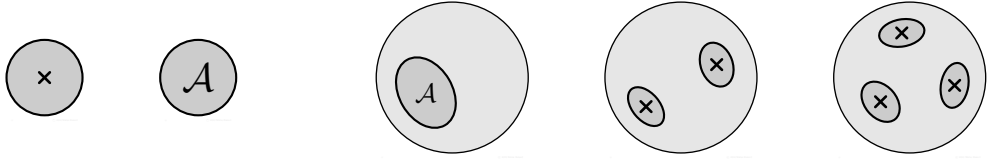
\centering
\includegraphicsbox[scale=\ifarxiv 1.0\else 0.8\fi]{FigCompStateChain.mps}
\qquad
\includegraphicsbox[scale=\ifarxiv 1.0\else 0.8\fi]{FigCompStateAmp.mps}
\qquad
\qquad
\includegraphicsbox[scale=\ifarxiv 1.0\else 0.8\fi]{FigCompCorr1.mps}
\qquad
\includegraphicsbox[scale=\ifarxiv 1.0\else 0.8\fi]{FigCompCorr2.mps}
\qquad
\includegraphicsbox[scale=\ifarxiv 1.0\else 0.8\fi]{FigCompCorr3.mps}
\caption{Large-$N\indup{c}$ topology representation 
of a single-trace local operator 
and a single-trace particle configuration (left two);
a cross corresponds to a puncture of the surface
serving as a source for charges. 
Quantum correlation functions insert a 
surface ending on the traces:
planar scattering amplitude (middle) 
as well as two- and three-point correlators (right two);
not shown are non-planar contributions 
where the surfaces have additional handles.}
\label{fig:StateCorr}
\end{figure}

Integrability and the Yangian algebra is 
tightly related to the 't Hooft planar limit \cite{'tHooft:1974jz}.
Let us therefore consider the large-$N\indup{c}$ expansion.
Local operators and particle configurations 
can be viewed as closed one-dimensional contours
such that each trace corresponds to one connected component;
let us refer to them as \emph{states},
see \figref{fig:StateCorr}.
Quantum correlation functions then span 
two-dimensional surfaces between the contours. 
The topology of these surfaces determines
the suppression in powers of $1/N\indup{c}$.
For example, a single-trace scattering amplitude as well as 
two- and three-point correlators of local operators
are displayed in \figref{fig:StateCorr}.
This is in line with the picture from the AdS/CFT 
dual string theory on $AdS_5\times S^5$,
where the surface is the string world sheet 
and its boundaries (or punctures) correspond to states. 

\begin{figure}
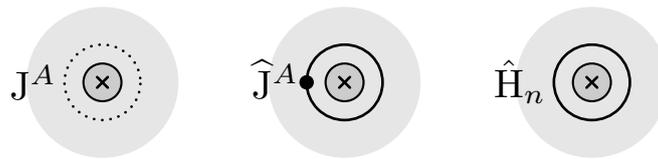
\centering
\includegraphicsbox[scale=\ifarxiv 1.0\else 0.8\fi]{FigCompStateConf.mps}
\qquad
\includegraphicsbox[scale=\ifarxiv 1.0\else 0.8\fi]{FigCompStateYang.mps}
\qquad
\includegraphicsbox[scale=\ifarxiv 1.0\else 0.8\fi]{FigCompStateHam.mps}
\caption{Action of generators on a single-trace state:
Superconformal $\gen{J}^A$,
Yangian $\genY{J}^A$ and local 
integrable Hamiltonians $\ham_n$.
Fat lines represent Wilson lines;
open loops have a marked base point, closed loops do not.
Dotted lines are plain integrals which can be broken up.}
\label{fig:GenState}
\end{figure}

Algebra generators are represented 
by Wilson loops of the Lax family of
flat connections on the world sheet \cite{Bena:2003wd}.
Thus we should represent a generator 
by a closed loop on the surface.
Due to flatness, the contour can be deformed smoothly.
The action on a state
corresponds to winding the loop around 
the trace, cf.\ \figref{fig:GenState}.

\begin{figure}
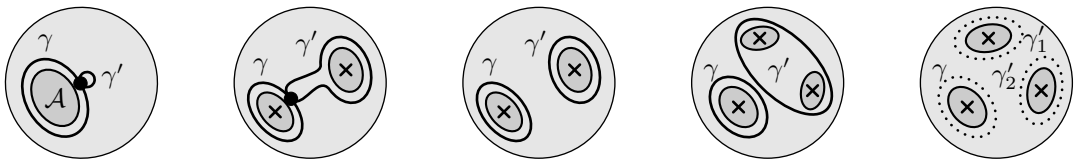
\centering
\includegraphicsbox[scale=\ifarxiv 1.0\else 0.8\fi]{FigCompCorr1Yang.mps}
\qquad
\includegraphicsbox[scale=\ifarxiv 1.0\else 0.8\fi]{FigCompCorr2Yang.mps}
\qquad
\includegraphicsbox[scale=\ifarxiv 1.0\else 0.8\fi]{FigCompCorr2Ham.mps}
\qquad
\includegraphicsbox[scale=\ifarxiv 1.0\else 0.8\fi]{FigCompCorr3Ham.mps}
\qquad
\includegraphicsbox[scale=\ifarxiv 1.0\else 0.8\fi]{FigCompCorr3Conf.mps}
\caption{Invariance conditions can be understood through
deforming the contours $\gamma\to\gamma'$ associated to generators on the surface.
Yangian invariance follows from shifting the contour around the surface
and shrinking it to a point (leftmost). 
For a two-point function Yangian invariance is broken because 
the marked base point cannot be moved (middle left);
the integrable charges are nevertheless conserved (middle).
For a three-point function even the integrable charges are
not conserved (middle right), but the conformal ones are
(rightmost).}
\label{fig:Invariance}
\end{figure}

Now we can consider invariance conditions for particular objects,
see \figref{fig:Invariance}: 
In \secref{sec:scattering} we have seen that (tree-level) planar single-trace amplitudes
are invariant under Yangian symmetry.
In our picture we should wind an open Wilson loop around the trace
of the particle configuration. We can unwind the Wilson loop on the disc 
ending on the trace without having to move the base point.
The Wilson loop then shrinks to a point implying invariance.
The unwinding would not be possible in the non-planar case 
of a disc with handles.

For a correlation function of two traces, e.g.\ a two-point function
of single-trace local operators, the Yangian action 
on the two states is not equivalent because the base point is different.
A closed Wilson loop can, however, be deformed from one trace to the other
in the planar case of an annulus connecting the traces.
This implies the integrability of 
the problem of planar anomalous dimensions.%
\footnote{For planar double-trace scattering amplitudes
integrability implies the existence of a tower of conserved charges.
It would be interesting to confirm them.}

Finally, a correlation function of three traces does not have conserved charges. 
Merely superconformal symmetry survives, because for these generators
the loop is an abelian contour integral which can be broken up into two pieces.

\subsection{Summary and Outlook}

In this paper we have reviewed Yangian symmetry 
which serves as an algebraic foundation of 
integrability in planar $\superN=4$ maximally supersymmetric gauge theory.

We have seen that the Yangian is capable of uniquely determining certain 
physical observables by purely algebraic means. 
Even more importantly, there exist methods to exploit the uniqueness
and obtain these observables very efficiently. 
Among them are the Bethe ansatz, spectral curves, 
asymptotic Bethe equations with L\"uscher corrections,
thermodynamic Bethe ansatz or Y-system and Gra\ss{}mannians. 
Applying them one can avoid highly complicated calculations in quantum field theory 
and arrive at the correct final result much faster.

Although these methods are already being applied reliably, 
it still remains to be understood \emph{why} they work. 
Why is planar $\superN=4$ SYM governed by a Yangian algebra
(technically as well as semantically)?
How does it lead to the above methods?
How is the algebra defined in the first place?
As we have seen, at leading perturbative order the Yangian follows precisely 
from the established framework of quantum algebra.
At higher loops the Yangian representation gets deformed,
and some of the well-known rules have to be dropped
in favour of new ones yet to be established. 

We have discussed two subjects where the 
Yangian algebra makes a prominent appearance: 
anomalous dimensions of local operators and
the spacetime scattering matrix. 
A third subject which was not discussed here 
is the worldsheet scattering matrix.
The Yangian relevant to that problem 
is not based on $\alg{psu}(2,2|4)$ but
only on the subalgebra $\alg{psu}(2|2)$. 
Although this is an exception case, 
several works have demonstrated that 
it can apparently be described by conventional 
quantum algebra methods, see the review \cite{Torrielli:2010aa}. 
Complete understanding of the smaller Yangian may 
eventually lead to clues for the full perturbative
Yangian for $\superN=4$ SYM.

\pdfbookmark[1]{\refname}{references}
\ifarxiv
\bibliographystyle{nb}
\bibliography{lev70}
\else
\bibliographystyle{ws-procs9x6}
\bibliography{gribov80}
\fi

\end{document}